# Protease-sensitive atelocollagen hydrogels promote healing in a diabetic wound model

Giuseppe Tronci,[1,2*] Jie Yin,[1,2] Roisin A. Holmes,[2] He Liang,[1,2] Stephen J. Russell,[1] David J. Wood[2]

[1] Nonwovens Research Group, School of Design, University of Leeds, Leeds, United Kingdom

[2] Biomaterials and Tissue Engineering Research Group, School of Dentistry, University of Leeds, Leeds, United Kingdom

**Table of contents entry**

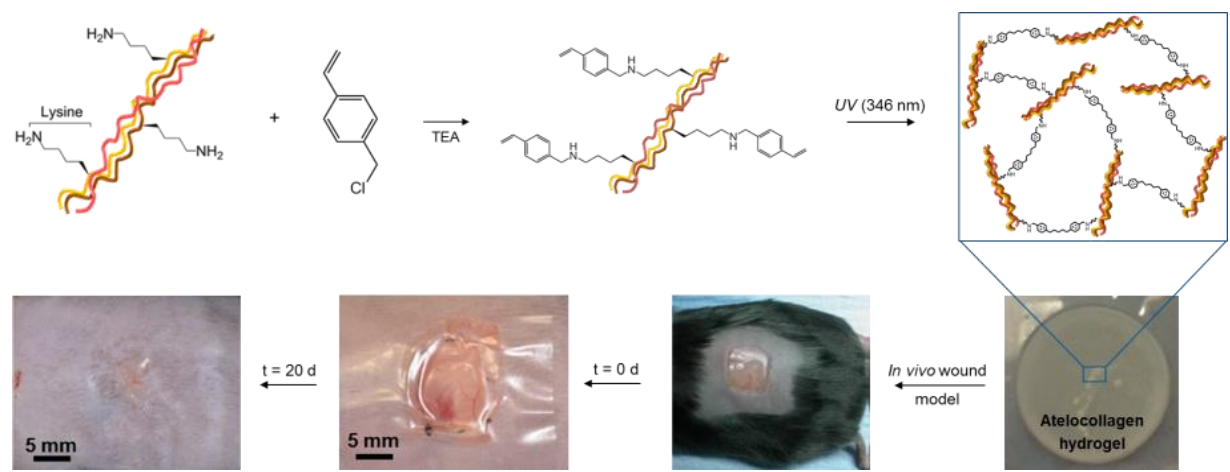

Protease-sensitive atelocollagen hydrogels were chemically designed to promote accelerated wound healing *in vivo* compared to a dressing gold standard.

[*] Corresponding author: Level 7 Wellcome Trust Brenner Building, St. James's University Hospital, University of Leeds, Leeds LS9 7TF, UK. E-mail address: g.tronci@leeds.ac.uk  (G. Tronci)

**Abstract**

The design of exudate-managing wound dressings is an established route to accelerated healing, although such design remains a challenge from material and manufacturing standpoints. Aiming towards the clinical translation of knowledge gained *in vitro* with highly-swollen rat tail collagen hydrogels, this study investigated the healing capability in a diabetic mouse wound model of telopeptide-free, protease-inhibiting collagen networks. 4-vinylbenzylation and UV irradiation of type I atelocollagen (AC) led to hydrogel networks with chemical and macroscopic properties comparable to previous collagen analogues, attributable to similar lysine content and dichroic properties. After 4 days *in vitro*, hydrogels induced nearly 50 RFU% reduction in matrix metalloproteinase (MMP)-9 activity, whilst showing less than 20 wt.-% weight loss. After 20 days *in vivo*, dry networks promoted 99% closure of 10×10 mm full thickness wounds and accelerated neo-dermal tissue formation compared to Mepilex®. This collagen system can be equipped with multiple, customisable properties and functions key to personalised chronic wound care.

**1. Introduction**

Chronic wounds in the form of venous leg ulcers (VLUs), diabetic foot ulcers (DFUs) and pressure ulcers (PUs) fail to repair in an orderly and timely self-healing process.[1,2] With increasing life expectancy and the associated occurrence of vascular diseases and type II diabetes, costs associated with chronic wound care represent a significant burden to healthcare systems worldwide and are expected to continue to rise. In the UK alone, ~ 650,000 patients are affected by such pathological conditions, resulting in a £3 billion annual cost to the National Health Service (NHS).[3] Additionally, chronic wounds are responsible for prolonged pain and morbidity in patients.

Dressing materials have been widely employed in the clinic for the treatment of chronic wounds.[4,5] In contrast to skin substitutes,[6] wound dressings are temporarily applied to the wound bed, in order to ensure a defined environment in terms of moisture (to minimise the risk of tissue maceration) and exudate management (to retain growth factors, MMPs and specific cells key to healing).[7,8] Furthermore, an ideal wound dressing should (i) provide thermal insulation and oxygen exchange; (ii) protect damaged tissue from secondary infections and bacterial contamination; (iii) display low adherence *in situ* to enable complete dressing removal without debris formation and integration with the host tissue; (iv) control activity of up-regulated MMPs, such as MMP-9,[9,10,11,12,13,14,15,16,17] in order to promote wound healing; (v) not induce any toxic response to tissue microenvironment. Although these requirements can be individually provided by many existing commercial dressings, such controlled multi-functionality is still challenging to accomplish in a single, soluble factor-free material system. Here, we report a synthetically processed, triple helix preserved collagen system that fulfils the above requirements and successfully leads to complete wound closure in diabetic mice. Moreover, the system can be customised to provide bespoke material architectures, i.e. hydrogels, fibres and fabrics.

As they are based on hydrophilic building blocks, hydrogels have been widely employed for the design of wound dressings;[18] their water content can be tuned in order to ensure defined levels of exudate *in situ*,[19,20,21] whilst the moist interface with the skin prevents dressing adherence and allows for easy removal. Additionally, hydrogels can be customised into fibres and fabrics, whereby the creation of both internal pores and fabric architecture offers advantages with regard to wound exudate management and material dressability.[22,23] Polyurethane[24] and methylcellulose[25] have been successfully employed for the development of wound dressing products, i.e. Mepilex® and Aquacel®, respectively. Particularly the methylcellulose-based materials can become significantly weaker in the wet-state, highlighting the narrow trade-off between water absorbency and hydrated mechanical properties.[26] Also, such materials, based on either synthetic or polysaccharide-based backbones, do not contain MMP cleavage sites, which are key for the dressing to act as a substrate for up-regulated proteases in order to stimulate wound healing.

In an effort to introduce MMP sensitivity, polymer networks have recently been synthesised as active systems to accelerate diabetic wound healing *in vivo*.[27] Protease-cleavable hyaluronic acid macromers have been electrospun to ensure controlled enzymatic degradability.[28] Polyethylene glycol (PEG) hydrogels have been synthesised to either provide sustained, enzymatically-responsive release of peptide drugs,[29] or provide real-time quantification of MMP activity[30] aiming to establish a relationship to cell viability following drug treatment.[31] Ultimately, Francesko et al. investigated polyphenolic compounds in multi-component collagen-based sponges as enzymatic inhibitors.[32]

Rather than synthetically introducing MMP sensitivity in covalent networks, we have recently reported the creation of inherently enzymatically degradable rat tail collagen hydrogels with retained triple helix organisation that have higher compressive modulus and comparable water uptake with respect to commercial dressings, e.g. Aquacel®.[33,34,35] Wet stable fibres have also been produced by sequential wet-spinning of collagen triple helices and post-spinning covalent crosslinking with an aromatic diacid.[36,37,38] In light of these promising reports, we investigated in this study whether functional systems could be realised by building molecular architectures defined by the biochemical attributes of the chronic wound microenvironment in order to achieve accelerated healing *in vivo*. We hypothesised that type I telopeptide-free, non-hydrolysed, medical grade AC could be used as highly-purified building block for the formation of largely hydrated systems with minimal antigenicity.[39,40] AC triple helices were chemically functionalised with 4-vinylbenzyl chloride (4VBC) and integrated in a covalent network via UV irradiation. 4VBC was employed due to the remarkable swelling ratio and hydrated compressive modulus previously observed in rat tail collagen materials.[41] Introduced 4VBC adducts were also hypothesised to mediate complexation with upregulated MMPs found in chronic wound exudates. Experiments were carried out *in vitro* to investigate the impact of either AC hydrogels or two commercial dressings, i.e. Aquacel® (Convatec) and Mepilex® (Mölnlycke Health Care), on MMP-9 activity. MMP-9 was selected since significantly elevated levels in MMP-9 have been reported in chronic compared with acute wound fluids, suggesting a direct correlation with the clinical wound state.[42] Further to the study *in vitro*, full thickness wounds were created in db/db diabetic mice as well-accepted animal models in wound healing research,[43,44] and treated with either AC hydrogel or above-mentioned wound dressing gold standards.

## 2. Materials and methods

### 2.1 Materials

An aqueous solution of medical grade type I AC from bovine corium (3 mg·ml$^{-1}$ in 10 mM HCl) was kindly provided by Collagen Solutions PLC. Rat tails were provided by the School of Dentistry, University of Leeds (UK) from which type I collagen was isolated in-house via acidic treatment of rat tail tendons. Phosphate-buffered saline (PBS, w/o $Ca^{2+}$ and $Mg^{2+}$ ions) was purchased from Lonza. All the other chemicals were purchased from Sigma-Aldrich.

### 2.2 Synthesis of atelo- and rat tail collagen hydrogels

Medical grade AC hydrogels were prepared by adopting previously-published protocols.[33,34,41] A solution of type I AC in 10 mM HCl (3 mg·ml$^{-1}$) was neutralized to pH 7.4. 1 wt.-% of Tween-20 (with respect to the initial solution volume) and 30 molar excess of both

4-vinylbenzyl chloride (4VBC) and triethylamine (TEA) (with respect to the molar content of free amino groups of collagen, i.e. ~ $3\times10^{-4}$ moles·g$^{-1}$) were added. After 24 h reaction, the reaction mixture was precipitated in 10 − 15 volume excess of pure ethanol and stirred overnight before centrifugation (10000 rpm, 30 min) and air-drying. The retrieved dry product was characterised via 2,4,6-trinitrobenzenesulfonic acid (TNBS) assay[45] and circular dichroism.[33,34] Hydrogels were prepared by dissolving reacted collagen in 10 mM HCl solution containing 1 wt.-% 2-hydroxy-1-[4-(2-hydroxyethoxy) phenyl]-2-methyl-1-propanone (I2959) photoinitiator. The suspension was cast in a 12-well plate (1 ml/well) and UV irradiated (346 nm, Spectroline) for 30 min from above and below. Formed hydrogels were thoroughly washed in distilled water and used for compression tests. Sample dehydration was carried out in ethanol-distilled water mixtures with increasing ethanol content. Rat tail collagen hydrogel controls were prepared in the same manner, except that the reaction with 4VBC was carried out with a 0.25 wt.-% collagen solution in 10 mM hydrochloric acid, as previously reported.[33,34]

## 2.3 Compression tests

PBS-equilibrated hydrogel discs (Ø: 18 mm; h: 7 mm) were compressed at room temperature with a compression rate of 3 mm·min$^{-1}$ (Instron ElectroPuls E3000). A 250 N load cell was operated up to complete sample compression.[34] Stress-strain curves were recorded and the compression modulus quantified as the slope of the plot linear region at 25-30% strain. Six replicates were employed for each composition. Data are presented as mean ± standard deviation (SD).

**2.4 Swelling ratio and gel content**

Dry atelo- and rat tail collagen networks of known mass ($m_d$) were individually incubated in PBS-containing vials at 25 °C for 24 hours. The swelling ratio (*SR*) was calculated according to Equation 1:

$$SR = \frac{m_s - m_d}{m_d} \times 100 \qquad \textbf{(Equation 1)}$$

where $m_s$ is the mass of PBS-equilibrated samples.

In addition to the swelling ratio, the gel content was determined as the overall portion of the covalent hydrogel network insoluble in 10 mM HCl solution.[46] Dry, freshly synthesised, atelo- and rat tail collagen networks of known weight ($m_d$) were equilibrated in 10 mM HCl solution for 24 hour. Resulting hydrogels were air dried and weighed. The gel content (*G*) was calculated according to equation 2:

$$G = \frac{m_1}{m_d} \times 100 \qquad \textbf{(Equation 2)}$$

where $m_1$ is the dry sample mass following incubation in 10 mM HCl. Five replicates were employed for both swelling and gel content tests. Data are presented as mean ± SD.

**2.5 MMP-9 activity and enzymatic degradation study**

Full length human proenzyme MMP-9 (PF038, Merck Millipore,UK) was activated in fluorescence buffer (50mM Tris-HCL, 10mM $CaCl_2$, 150mM NaCl and 0.05% Brij-35, pH 7.5, Sigma-Aldrich) in the presence of p-aminophenylmercuric acetate (Sigma-Aldrich) at 37°C for 2 hours, according to manufacturer's specifications. Samples of known mass ($m_d$: 3-5 mg, n=4) of dry AC networks and gold standard dressings, i.e. Mepilex® (Mölnlycke Health Care) and Aquacel® (Convatec), were individually placed in a 24-well plate and incubated with the activated MMP-9 solution (100 ng·mL$^{-1}$, 1.5 mL) in an orbital shaker (160 rpm, MaxQ mini 4450) at 37°C. The concentration of MMP-9 was selected according to MMP-9 levels found in chronic wound fluids.[42] Sample-free solution controls of either activated or non-activated MMP-9 were also included. After 4 days, the MMP-9 activity of the supernatant was determined via a standard commercial fluorometric assay (ab112146, Abcam, UK); whilst, 4-day incubated samples were collected, washed in an increasing distilled water/ethanol series (0, 25, 50, 80, 100 vol.-% ethanol) and air-dried. The mass of retrieved, air-dried samples ($m_4$) was measured and the relative mass ($\mu_{rel}$) calculated according to Equation 3, as described previously:[38]

$$\mu_{rel} = \frac{m_4}{m_d} \times 100 \tag{Equation 3}$$

Statistical analysis was carried out using OriginPro 8.5.1. Data normality was confirmed via the Shapiro-Wilk test. Significance of difference was analysed using one-way ANOVA with Bonferroni test. A $p$ value of less than 0.05 was considered significant. Data are presented as mean ± SD.

## 2.6 *In vivo* study

### 2.6.1 Animals

Forty-one diabetic, male, 10-week old mice (BKS.Cg-m Dock7$^m$ +/+ Lepr$^{db}$ /J, The Jackson Laboratory, USA) were housed in groups of two animals according to Home Office regulations and specific requirements for diabetic animals. Prior to experimentation, they were housed for a period of more than 7 days without disturbance, other than to refresh their bedding and to replenish their food and water provisions. After experimental wounding, animals were housed in individual cages (cage size 500 cm$^2$ with sawdust bedding, changed three times per week) at 23°C with 12-hour light/dark cycles. Animals were provided with food (Standard Rodent Diet) and water *ad libitum.* All *in vivo* procedures were carried out in a

Home Office licensed establishment under Home Office Licences (PCD: 50/2505; PPL: 40/3614; PIL: IBCEFDF55; PIL: I34817249).

**2.6.2 Creation of full-thickness experimental wounds, treatment and monitoring**

Following randomisation into four experimental groups, animals were anaesthetised on day 0 (isofluorane & air) and the dorsa of animal groups 1-3 shaved and cleaned with saline-soaked gauze. A single standardised full-thickness wound (10 × 10 mm) was created in the left dorsal flank skin of each experimental animal. The wound site was hydrated with 25 µl of sterile saline prior to application of either a commercial dressing or a dry collagen network. Wounds of animal group 1 (n=11) received gamma-sterilised AC networks placed in such a manner as to overlay the wound margins. The upper surface of the collagen network was hydrated with 25 µl of sterile saline following application *in situ*. Wounds in groups 2 and 3 (n=10) received either a commercial polyurethane (Mepilex®, Mölnlycke Health Care) or carboxymethyl cellulose dressing (Aquacel®, Convatec), respectively, so that the dressings overlaid the wound margins by ~5 mm in all directions (Figure S1). All wounds of groups 1-3 also received a transparent polyurethane adhesive film (Bioclusive™, Systagenix Wound Management) in order to secure previously-applied dressings. Control wounds (n=10) were treated with Bioclusive™ (Systagenix Wound Management), only. On post-wounding days 4, 8, 12 and 16, all animals were re-anaesthetised and the outer film dressing and any free debris removed. Wounds were gently cleaned using saline-soaked sterile gauze, blotted dry with sterile gauze and digitally photographed. Each wound was scored by two independent observers, as to whether it was displaying initiation of neo-dermal tissue generation activity,[44] and the percentage of responsive wounds compared between experimental groups. Following this, fresh, sterile samples of collagen networks, commercial dressings and secondary adhesive dressings were then applied. Following all anaesthetic events, animals were placed in a warm environment and were monitored until they had fully recovered from the surgical procedure. All animals received appropriate analgesia (buprenorphine) after surgery and additional analgesics as required.

At day 20, wounds of all animal groups were cleaned, assessed and digitally photographed. Animals received an intraperitoneal injection (30 µg·g$^{-1}$ body weight) of 5-bromo-2′-deoxyuridine (BrdU, Sigma) one hour prior to termination, in order to facilitate future detection of proliferating cells by immuno-histochemical analysis of tissue sections. Wounds and marginal tissues were harvested and processed for histological investigation.

**2.6.3 Image analysis of wound closure**

Image Pro image analysis software (version 4.1.0.0, Media Cybernetics, USA) was used to calculate wound closure from wound images over time in each of the experimental groups. At selected time points, the open area was measured for each wound and wound closure expressed as percentage ratio between post-wounding open area and initial area.

**2.6.4 Harvesting and processing of wound tissues**

On post-wounding day 20, all animals were humanely sacrificed by $CO_2$ asphyxiation (confirmed by cervical dislocation). Wounds and surrounding normal tissue were excised and fixed in 10% (v/v) neutral buffered formalin (Sigma, UK) for histological assessment. Excised tissue was sandwiched between two pieces of foam, prior to being placed in fixative, to reduce the extent of tissue curling. Fixed specimens were trimmed and bisected, generating two half wounds per site. Both halves were processed and embedded in paraffin wax. Specimens were orientated in such a fashion as to ensure that appropriate transverse sections of the wound could be taken.

**2.6.5 Histological evaluations**

Wax-embedded tissues from post-wounding day 20 were sectioned at 6 µm thickness and stained with haematoxylin and eosin (H&E). The histological wound width and the granulation tissue depth were quantitatively assessed using Image Pro image analysis software. Wound width measurements were taken through the centre of each wound.[44] Cranio-caudal contraction was expressed as the percentage of the histological wound width measured at day 20 with respect to the central wound width calculated digitally at day 0 as an average of three separate measurements.

### 2.6.6 Statistical analysis

Data were tested for normality using the Shapiro-Wilk test. Fisher's exact test (for proportionate data) was carried out on a two-sample basis to analyse the impact of treatment on initiation of neo-dermal tissue repair activity. Non-parametric analysis (Kruskal-Wallis followed by ad hoc two-sample Mann Whitney U-test) was used to test the significance of any inter-group differences in wound closure, contraction and re-epithelialisation as well as granulation tissue depth and cranio-caudal contraction.

## 3. Results and discussion

The creation of a single-material AC system is presented, whereby macroscopic properties and functions can be controlled by structural parameters introduced at the molecular level in order to fulfil the complex requirements of the chronic wound microenvironment. The system is built from a defined molecular architecture of AC triple helices that are covalently functionalised and integrated in a covalent network following photo-activation (Scheme 1). The AC network swells upon contact with the wound exudate, forming hydrogels with superior exudate uptake and compression modulus with respect to two wound dressings routinely employed in the clinic, whilst providing additional regulation of upregulated MMP-9 activity. In the following, results on AC structural organisation and respective 4VBC-mediated functionalisation are presented and discussed in relation to the ones of type I rat tail collagen, as typical source employed for the formation of collagen-based materials. Subsequently, the attention moves to the characterisation of AC hydrogels *in vitro* and *in vivo*, aiming to explore their potential applicability in chronic wound care. Nomenclature used to present the results is as follows: type I AC networks are coded as '4VBC*', where '4VBC' refers to the functionalised precursor synthesised via reaction of AC with 30 molar excess of 4VBC with respect to the primary amino groups and amino termini of collagen, whilst '*' indicates the resulting covalent network obtained via UV irradiation.

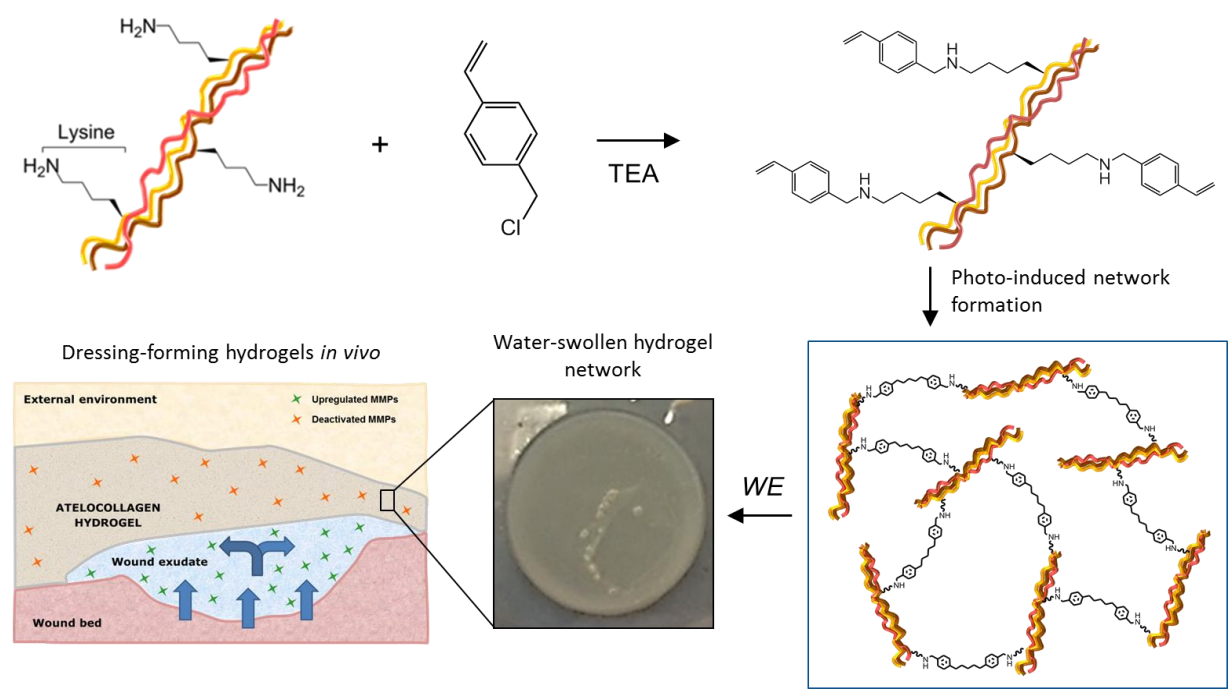

**Scheme 1.** Research strategy undertaken to realise AC hydrogels with superior wound exudate and MMP-9 management capabilities compared to current leading wound dressings. AC triple helices were functionalised with 4VBC adducts via lysine-initiated substitution reaction. UV irradiation of AC precursor solutions resulted in the creation of 4VBC-covalently crosslinked networks of retained AC triple helices. In contact with wound exudate (*WE*), networks swell resulting in dressing-forming hydrogels at the macroscopic scale. When applied to hard-to-heal wounds, hydrogels display high exudate uptake *and* compression modulus, and control the activity of exudate-carried upregulated MMPs via complexation with 4VBC adducts.

### 3.1 Synthesis of type I AC networks

A molar content of primary amino groups of about $3\times10^{-4}$ mol·g$^{-1}$ was measured via TNBS on pristine type I AC, in agreement with previously-reported values.[33-35,37,41] Likewise, far-UV CD analysis indicated typical dichroic spectral patterns of type I collagen, i.e. a positive triple helix-related maximum absorption band at 220-225 nm, and a negative minimum absorption band around 195-200 nm, associated with the presence of polyproline II helices (Figure S2). Quantitative analysis of the AC CD spectrum resulted in a magnitude ratio between positive and negative peak intensities (*RPN*) of 0.13, in line with previous findings.[33,34] Reaction of type I AC with 4VBC was carried out prior to UV irradiation in the presence of I2959 photoinitiator and resulted in the formation of covalent hydrogel networks (*G*: 97 ± 4 wt.-%, Table 1). 4VBC-functionalised precursors displayed 45 mol.-% averaged functionalisation of primary amino groups, whilst Far-UV CD confirmed retention of triple helices (*RPN*: 0.14, Table 1). PBS-equilibrated hydrogels exhibited a swelling ratio of about 2000 wt.-% and a compressive modulus of about 80 kPa, confirming comparable macroscopic properties with respect to rat tail collagen-based analogues.

**Table 1.** Chemical, structural and hydrogel properties of 4VBC-functionalised AC networks. (*F*): Degree of functionalisation, (*RPN*): CD magnitude ratio between positive and negative peak intensities, (*SR*): swelling ratio following equilibration with PBS, (*G*): gel content, ($E_c$): compressive modulus. Data is presented as mean ± SD.

| Sample ID | *F* /mol.-% | *RPN* | *G* /wt.-% | *SR* /wt.-% | $E_c$/kPa |
|---|---|---|---|---|---|
| 4VBC* | 45 ± 4 | 0.14 | 97 ± 4 | 2065 ± 191 | 84 ± 17 |
|  | 44 ± 1 [(a)] | 0.12 [(a)] | 100 ± 0 [(a)] | 1900 ± 200 [(a)] | 81 ± 9 [(a)] |

[(a)] Rat tail collagen control.

These values proved to be higher than the ones displayed by two gold wound dressing standards, i.e. Aquacel® (*SR*: 1759 ± 107 wt.-%; $E_c$: 34 ± 18 kPa) and Mepilex® (*SR*: 447 ± 53 wt.-%).

### 3.2 Protease sensitivity and degradability *in vitro*

Samples of 4VBC*, Mepilex® and Aquacel® were incubated in a buffer solution containing activated MMP-9. Figure 1 displays the MMP-9 activity of sample-treated supernatants relative to the one of sample-free supernatants (containing activated MMP-9), as well as the relative mass ($\mu_{rel}$) of collected samples, after 4 days (as the typical application time of a chronic wound dressing). The MMP-9 activity was found to be decreased to 51±5 RFU% when the respective solution was in contact with samples of 4VBC*. In contrast, a non-statistically significant variation in MMP-9 activity was observed following application of either Mepilex® or Aquacel®.

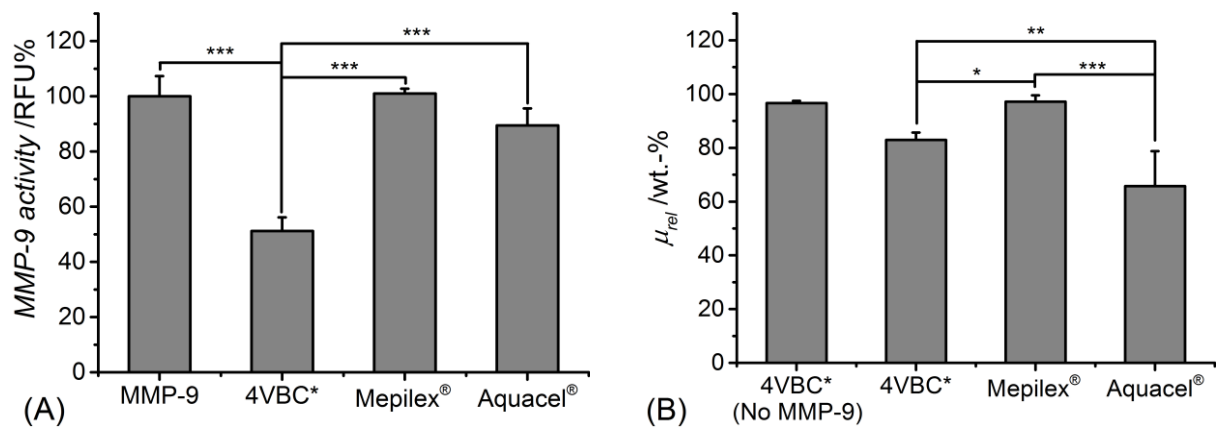

**Figure 1.** (A) MMP-9 activity measured in the supernatant following 4-day incubation with either no sample (MMP-9) or sample of 4VBC*, Mepilex® or Aquacel®. Data are expressed as relative percentages with respect to the activity of untreated MMP-9 solution. (B): Relative sample mass ($\mu_{rel}$) of either 4VBC*, Mepilex® or Aquacel® following 4-day incubation in an aqueous solution with or without activated MMP-9. *: $p < 0.05$; **: $p < 0.01$; ***: $p < 0.001$.

In the presence of MMP-9, samples of 4VBC* displayed more than 80 wt.-% averaged relative mass ($\mu_{rel}$: 83±2 wt.-%), a value found to be between the one measured in samples of Mepilex® ($\mu_{rel}$: 97±2 wt.-%) and Aquacel® ($\mu_{rel}$: 66±13 wt.-%). On the other hand, control experiments in MMP-9–free media revealed minimal weight loss in the case of 4VBC* ($\mu_{rel}$: 97±1 wt.-%) and Mepilex® ($\mu_{rel}$: 99±1 wt.-%, data not shown), whilst similar trends were observed in the case of Aquacel® ($\mu_{rel}$: 70±4 wt.-%, data not shown).

### 3.3 Macroscopic wound assessment

A schematic overview of the overall strategy pursued to monitor the healing process is shown in Figure 2. Macroscopic closure was expressed as the ratio between the post-wounding open and original areas, whereby wound dimensions were quantified digitally (Figure 2, A and B).

The extent of granulation tissue formation, re-epithelialisation and contraction to wound closure was determined either macroscopically or histologically. Wound contracted and re-epithelialised areas were identified from time-specific digital macrographs (Figure 2, B), whilst granulation tissue depth and wound width were quantified via histological H&E tissue sections at day 20 post-wounding (Figure 2, C). All experimental groups displayed nearly-complete closure of open wound areas at the end of the study *in vivo* (Figure 3, I-K), in contrast to the control group showing minimal dimensional changes (Figure 3, L).

Among the three dressing groups, a major reduction of open wound area was observed after 8 days in animals treated with both 4VBC$^*$ and Aquacel$^®$. All mice were found to tolerate the application of the AC hydrogel and the two commercial dressings, although some erythema was noted in the skin surrounding the wounds (4VBC$^*$: n=3, Mepilex$^®$: n=4, Aquacel$^®$: n=2). This may be attributed to the repeated application and removal of the adherent film Bioclusive$^{TM}$. Other than that, hydrogel 4VBC$^*$ remained intact on the wound surface during the 4-d application time window and could be easily removed. In contrast, samples of Aquacel$^®$ were found in different locations on day 4 with respect to the underlying wound position.

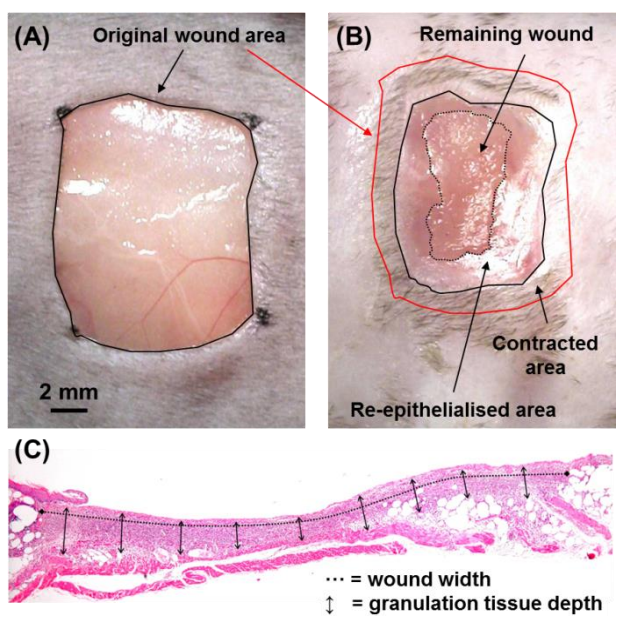

**Figure 2.** Strategy adopted to monitor the wound healing process over a 20-d in vivo study. (A-B): the wound area was quantified via equally-scaled digital macrographs taken at specific time points. Wounds were optically assessed in terms of neo-dermal tissue generation, contraction and re-epithelialisation. (C): Histology sections were obtained at the end of the study so that wound width and granulation tissue depth were quantified.

### 3.4 Temporal profiles of wound closure, re-epithelialisation and contraction

Post-wounding digital macrographs were analysed to determine (i) the percentage of wounds showing initiation of neo-tissue formation and (ii) the temporal profiles of wound closure, contraction and re-epithelialisation (Figure 4). In comparison to the adhesive-treated control group, significant improvements ($p \leq 0.011$) in both neo-dermal tissue generation activity and closure were observed in dressing-treated wounds at all assessment time points (Figure 4, A and B) irrespective of the dressing type. More than 80% response was identified at day 4 in wounds in receipt of AC hydrogel, a percentage found to be between the one recorded in Aquacel®- (100%) and Mepilex®- (60%) treated experimental groups (Figure 4, A); whilst no responding wound was found at this assessment point in the adhesive-treated control group. By day 8, neo-dermal tissue formation occurred in all wounds in receipt of both 4VBC* and Aquacel® dressings, whilst application of Mepilex® dressing promoted neo-tissue stimulation in 80% wounds, in stark contrast with the control group (10%). Likewise, temporal profiles of wound closure (Figure 4, B) indicated a significantly reduced open area in either 4VBC*- and Aquacel®-, compared to Mepilex®-, treated wounds ($p=0.024$). From day 12 onwards, all experimental groups confirmed initiation of tissue formation, whilst this was still minimal in the control group. Significantly increased wound closure was observed following application of Aquacel®, compared to Mepilex®, dressing on day 12 ($p = 0.043$), whilst comparable results were observed when sample 4VBC* was employed (Figure 4, B). Other than neo-tissue initiation and wound closure, re-epithelialisation and contraction temporal profiles (Figure 4, C and D) showed that Mepilex® dressing induced the highest level of re epithelialisation (day 20: $34 \pm 15$ %) at all selected post wounding time points, followed by 4VBC* hydrogel (day 20: $27 \pm 10$ %) and Aquacel® dressing (day 20: $22 \pm 5$ %), whilst the opposite trend became apparent with regard to the contraction profile (day 20: 4VBC*: $70 \pm 9$ %; Mepilex®: $66 \pm 14$ %; Aquacel®: $78 \pm 5$ %).

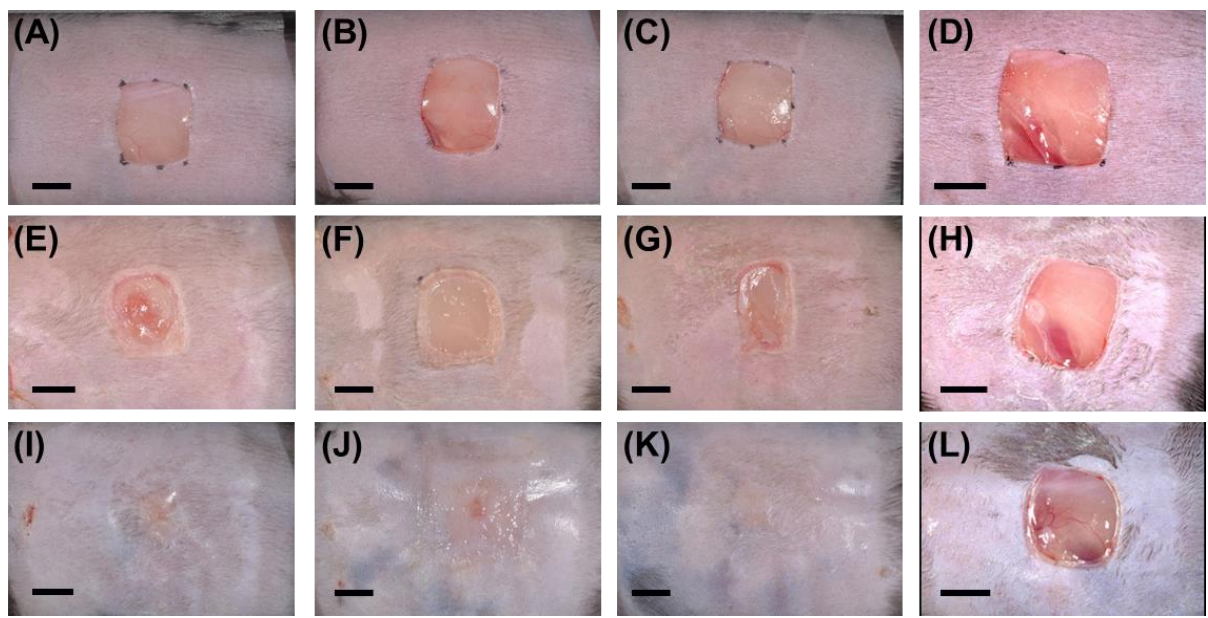

**Figure 3.** Digital macrographs of the wound site (10 × 10 mm) at day 0 (A-D), 8 (E-H) and 20 (I-L), following application of dry AC hydrogel (4VBC* (A, E, I)), commercial polyurethane foam dressings (Mepilex® (B, F, J)), commercial cellulose-based dressings (Aquacel® (C, G, K)), and commercial adhesive controls (Bioclusive™ (D, H, L)). The scale bar is 5 mm.

Levels of re-epithelisation were found to increase up to day 12 post-wounding, after which they tended to plateau or reduce due to the on-going contraction of the wounds. From day 16 onwards, wounds in receipt of AC hydrogel showed significantly greater re-epithelialisation ($p \leq 0.043$) with respect to wounds in receipt of Aquacel® dressing; and a similar situation was apparent when comparing groups treated with Mepilex® and Aquacel® dressings from day 12 onwards ($p \leq 0.029$). The latter dressing promoted significantly higher wound contraction with respect to 4VBC* hydrogel from day 12 onwards ($p \leq 0.024$), and similar results were obtained when comparing the former with the Mepilex® dressing group in the same time window ($p \leq 0.011$).

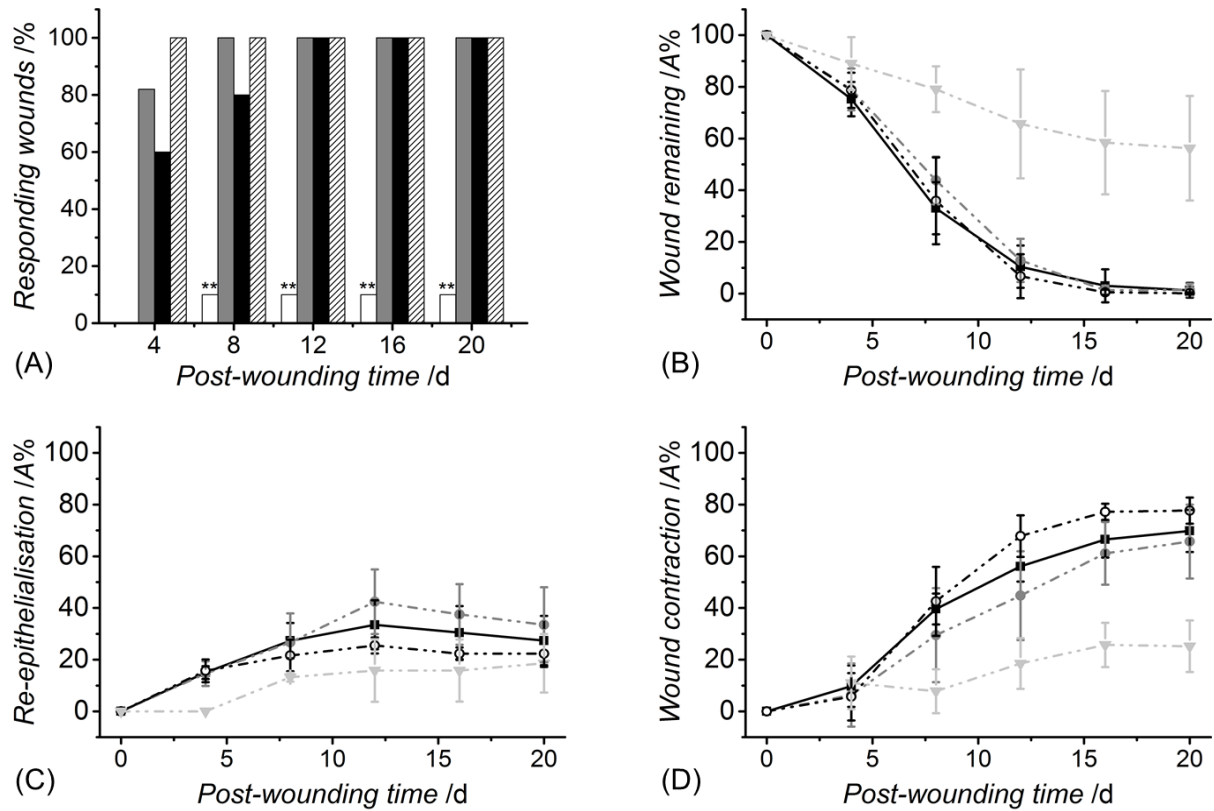

**Figure 4.** (A): Percentage of responding wounds showing initiation of neo-dermal tissue formation at selected post-wounding time points following application of commercial adhesive controls (Bioclusive®, white columns), AC hydrogels (4VBC*, grey columns), commercial polyurethane foam dressings (Mepilex®, black columns), and commercial cellulose-based dressings (Aquacel®, patterned columns). **: $p < 0.01$. At day 4, no responding wound was found in the control group, so that no column is reported at this assessment point. (B-D): Temporal profiles of remaining wound area (B), wound re-epithelialisation (C) and wound contraction (D) determined from post-wounding digital macrographs upon receipt of AC hydrogels (4VBC*, –■–), commercial polyurethane foam dressings (Mepilex®, ···●···), commercial cellulose-based dressings (Aquacel®, ···○···) and commercial adhesive controls (Bioclusive™, —▼—). Curves are guidelines to the eye. Data are presented as the mean ± SD.

### 3.5 Histological analysis of cranio-caudal contraction and granulation tissue depth

Typical low power wound histology images for all treatment groups at day 20 post-wounding are shown in Figure 5. Granulation tissue depth (GTD) was significantly increased in wounds with either 4VBC* (Figure 5, A), Mepilex® (Figure 5, B) or Aquacel® (Figure 5, C) dressing, compared to control wounds (Figure 5, D), confirming previous trends obtained at the macro-scale. Comparable GTD levels were found upon application of AC and cellulose-based materials, with marginally reduced levels in wounds treated with Mepilex® dressing (Figure 6).

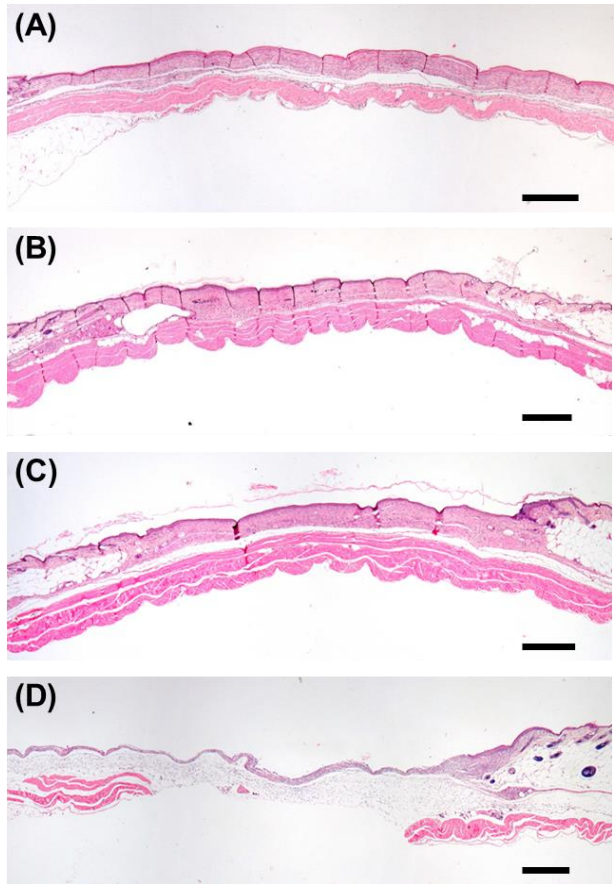

**Figure 5.** Low power images of typical histological sections related to 20-d wounds treated with either AC hydrogels (4VBC[*], A), commercial polyurethane foam dressings (Mepilex®, B), commercial cellulose-based dressings (Aquacel®, C) or commercial adhesive controls (Bioclusive[TM], D). The scale bar is 500 μm.

Further to GTD, cranio-caudal contraction (CCC) was quantified in order to express the reduction in histology-based central wound width at the time of tissue harvesting (Figure 2, C) relative to that at the time of injury. CCC values (Figure 6) proved to be comparable to macroscopic values of wound contraction measured at day 20 (Figure 4, C). Wounds in receipt of either 4VBC[*], Mepilex® or Aquacel® dressings exhibited significantly increased CCC with respect to control wounds (p = 0.000). Samples of Aquacel® promoted the greatest levels of CCC, followed by the AC hydrogel and ultimately by samples of Mepilex®, although no significant differences were observed.

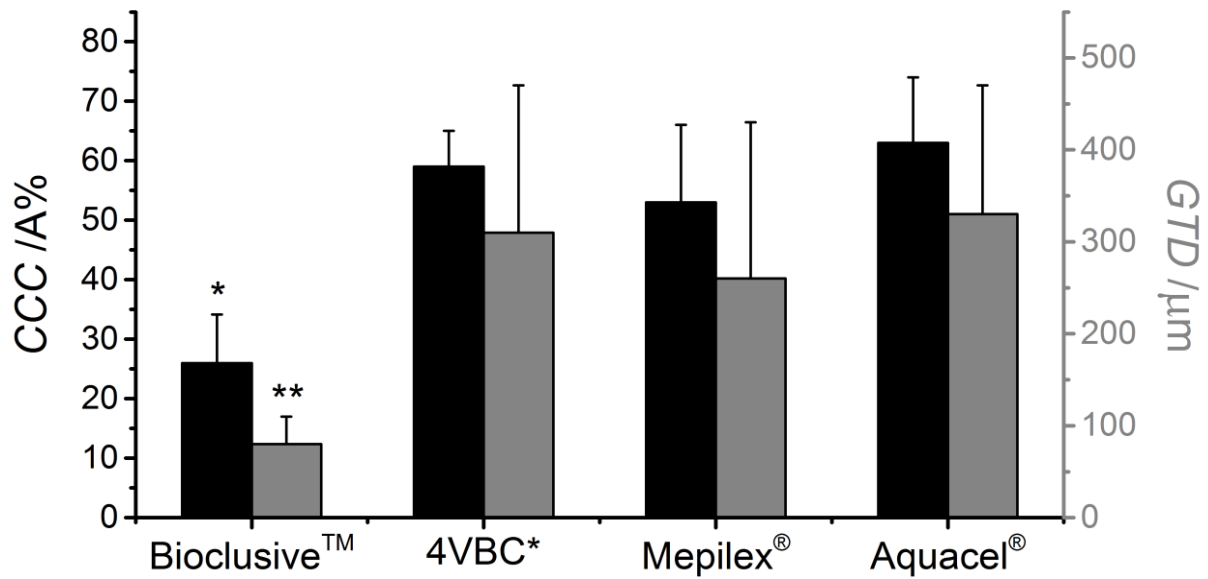

**Figure 6.** Histology-based cranio-caudal contraction (*CCC*, black columns, left-handed y axis) and granulation tissue depth (*GTD*, grey columns, right-handed y axis). *CCC* was calculated as the ratio between the wound width measured on 20-d H&E histology sections and the wound width measured on day 0 digital macrographs. '*' and '**' ($p = 0.000$) indicate means significantly different from the means of the other groups (non-parametric Mann Whitney U-test).

### 3.6 Discussion

Several natural compounds have been proposed for the treatment of hard-to-heal wounds, such as Manuka honey,[23] bee venom,[47] propolis[48] and whey protein,[49,50] in light of their antibacterial and antioxidant activity, although their potential effect in wound healing is still matter of current research. Aiming to develop intelligent wound dressings capable to accelerate the healing of either delayed or impaired wounds, collagen has recently gained a great deal of attention. Several collagen-containing systems have been approved for clinical use, including Promogran® (Systagenix),[51] a lyophilized collagen-oxidised regenerated cellulose matrix that gels on contact with wound exudate, and Biostep™ (Smith & Nephew, Inc),[52] a blend consisting of collagen, sodium alginate, carboxyl methylcellulose, and ethylenediaminetetraacetic acid (EDTA). In hydrated conditions, these dressings present documented form-instability in biological fluids,[53] in comparison with non-collagen-based standards of care, e.g. Aquacel® and Mepilex® (both used in this study), resulting in handling difficulties, debris formation and bacterial contamination *in vivo*. Such poor behaviour is partially explained by the fact that collagen has been reported to lose its native triple helix architecture when blended with other polymers (potentially requiring the use of organic solvents) and processed with state of the art manufacturing methods,[54] resulting in hydrolysed, denatured derivatives.[37,55] In light of these challenges, alternative biomaterials have therefore been proposed and explored either *in vitro* or pre-clinically for the treatment of burn injuries and impaired wounds, including copper-doped glass microfibres inducing angiogenesis and wound closure,[56] nanofibrous peptidic hydrogels,[57] protease-cleavable polymer systems,[28-31] as well as hydrogels containing protease-inhibiting soluble factors[13,32] or cells,[58] often requiring considerations with regard to activity and release of payload as well as customisation of material format.

The use of individual, ECM-derived biopolymers is a promising strategy for chronic wound care aiming to create wound-customised systems capable to inherently regulate MMP activity and manage wound exudate via structural parameters introduced at the molecular level. Here, the selection of purified, epitope-free sources[59] and the development of multiscale approaches[60] enabling bespoke structure-property-function relationships and material formats are key tasks to enable successful translation to clinical use. The results of this study provide a systematic and comprehensive investigation on the design of a polymer-, soluble factor- and cell-free AC system with controlled MMP-9 sensitivity and enzymatic degradability, displaying retained triple helices, and capable of promoting nearly 99% closure in diabetic wound models, comparable to established gold standards of care.

To investigate wound healing capability, medical grade non-hydrolysed AC was selected as a pristine backbone and compared to rat tail collagen. The influence of the collagen source on the physical and structural properties of resulting materials is known[59] and was therefore addressed by characterising both native as well as functionalised and crosslinked states. Especially for applications *in vivo*, the selection of antigen-free sources is a key to the design of medical collagen products, whilst the presence of non-collagenous, potentially immunogenic impurities, e.g. deriving from the material synthesis, should also be considered. Antigenic determinants of collagen can be found in the (triple) helical regions, with variations in the amino acid sequences not exceeding more than a few percent between mammalian species.[39,40] A far greater degree of variability is found in the non-helical terminal regions, i.e. telo-peptides, with up to half of the amino acid residues in these regions exhibiting interspecies variation.[61] Although a systematic immune-toxicity evaluation was not the focus of this study, two strategies were adapted to ensure the formation of collagen materials with minimal antigenicity and immunogenicity: (1) selection of a telopeptide-free collagen source compatible with the synthesis of 4VBC-functionalised collagen networks, and (2) intensive material washing following each synthetic step aiming to accomplish complete removal of non-collagenous impurities. Selected type I AC was obtained via pepsin-mediated extraction of bovine corium, resulting in a highly purified telopeptide-free backbone with similar lysine content and dichroic properties to our previously-used in-house extracted collagen (Table 1).

Telopeptide-free hydrogels were successfully synthesised with enhanced swelling ratio, compressive modulus (Table 1) and enzymatic degradability (Figure 1), compared to both Aquacel® and Mepilex®. Samples proved to display less than 20 wt.-% mass loss in both enzymatic and hydrolytic conditions, whilst still inducing nearly 50 RFU% reduction in MMP-9 activity *in vitro*, in contrast to both commercial dressings. The synthesis of the covalent network was a key to provide AC hydrogels with statistically decreased degradability, with respect to the case of Aquacel® (as based on non-crosslinked, water-soluble carboxymethyl cellulose). Covalently-bound electron-rich aromatic adducts in hydrogels 4VBC* were expected to be responsible for the observed inhibition of MMP-9 activity via chelation with zinc sites of active MMPs, minimising the need of soluble, protease-chelating factors, e.g. EDTA, in the collagen system. The combination of these swelling, degradation and compression properties together with the preserved triple helix organisation at the molecular level proved to be crucial in order to promote accelerated wound healing in the hydrogel- with respect to Mepilex®-treated wounds; after 20 days post-wounding and in contrast to control wounds, similar levels of wound contraction, re-epithelialisation (Figure 4), cranio-caudal contraction and granulation tissue depth (Figure 6) were observed in wounds in receipt of 4VBC* with respect to wounds treated with both dressings. The high hydrogel swellability in physiological conditions promoted sequestration and deactivation of up-regulated proteases, so that proteases could be diverted from the neo-tissue to the dressing and healing accelerated. In line with gravimetric enzymatic degradation data, AC dressings exhibited minimal macroscopic change following a 4-day application *in vivo*, so that they could be easily located at and removed from the wound bed with no sample break. After 20 days post-wounding, telopeptide-free hydrogels were found to induce both wound contraction and re-epithelialisation. These events are essential for successful skin wound healing, in order to quickly re-establish barrier function by migration of keratinocytes in the direction of the injury and minimise risks of environment-triggered insults.[62] Wound and histological analyses (Figure 2 and 4) confirmed that the regeneration of the epidermis was almost complete for hydrogel-treated wounds, suggesting the presence of healthy epithelial cells on the top of the injured tissue, whilst new epidermis appeared as thick as that of healthy skin wounds.

## 4. Conclusions

The present study provided new information on the capacity of a functionalised collagen network to stimulate healing of full-thickness wounds in diabetic mice, avoiding the use of soluble factors, encapsulated cells or multi-compartment systems. A medical grade, AC source was identified as telopeptide-free analogue of in-house extracted rat tail collagen, enabling the formation of purified hydrogels with accelerated neo-dermal tissue initiation capability in comparison to a clinical polyurethane dressing, i.e. Mepilex®. The retention of native triple helices in and high swellability of collagen networks enabled the development of a hydrated wound bed *in situ*, whereby controlled enzymatic degradability and inherent control in MMP activity could be accomplished. When used to treat full thickness wounds in diabetic mice, hydrogels showed a significantly better capacity to stimulate wound closure and re-epithelialisation than the untreated wounds (control), whilst similar healing levels were observed when either Aquacel® or Mepilex® gold-standard dressings were applied. The versatility of this collagen system is expected to promote the design of advanced, protease-degradable wound dressings with systematically adjusted molecular, microscopic and macroscopic architecture to enable superior wound exudate management capability and healing performance *in vivo*.

**Acknowledgments**

The authors thank the Clothworkers Centre for Textile Materials Innovation for Healthcare, EPSRC Centre for Innovative Manufacturing in Medical Devices (MeDe Innovation), EPSRC MeDe Innovation Fresh Ideas Fund and the University of Leeds Medical Technologies IKC for financial support. Dr. Jeff Hart and Andrea Bell (Cica Biomedical Ltd., UK) are gratefully acknowledged for carrying out the in vivo work. The authors thank Dr Graeme Howling and Martin Fuller for their support with the study *in vivo* and TEM, respectively.

**Supporting information**

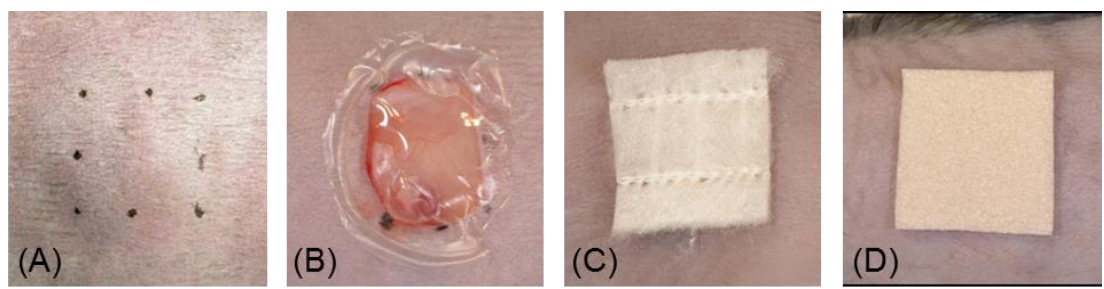

**Figure S1.** (A-D): digital macrographs taken during in vivo surgery. (A): The mouse dorsum was shaved and incised to create a 10×10 mm wound (A). (B-D): AC hydrogel (B) and two commercial dressings, i.e. Aquacel® (C) and Mepilex® (D), were applied on the top of the wound. Samples were replaced every four days over a 20-d time window.

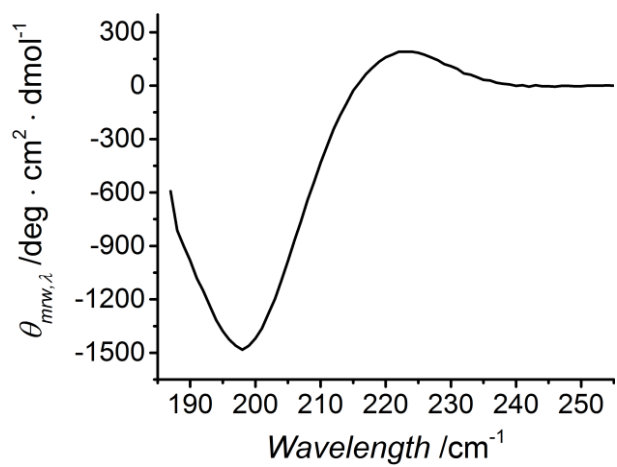

**Figure S2.** Far-UV CD spectrum of AC solution in 10 mM HCl. 195-200 and 220-225 nm peaks are clearly detected, confirming the presence of collagen polyproline II and triple helices, respectively.

# References

1 W.A Sarhan, H.M.E. Azzazy, I.M. El-Sherbiny, *ACS Appl. Mater. Interfaces*, 2016, **8**, 6379-6390.

2 F. Werdinemail, M. Tenenhaus, H.-O. Rennekampff, *The Lancet*, 2008, **372**, 1860-1862.

3 A. McLister, J. McHugh, J. Cundell, J. Davis, *Adv Mater* 2016, **28**, 5732-5737.

4 K. Vowden, P. Vowden, *Surgery*, 2014, **32**, 462-467.

5 J.L. Schiefer, R. Rath, M. Held, W. Petersen, J.O. Werner, H.E. Schaller, A. Rahmanian-Schwarz, *Adv Skin Wound Care*, 2016, **29**, 73-78.

6 H. Lagus, M. Sarlomo-Rikala, T. Böhling, J. Vuola, *Burns*, 2013, **39**, 1577-1587.

7 J.S. Boateng, K.H. Matthews, H.N.E. Stevens, G.M. Eccleston, *J Pharm Sci*, 2008, **97**, 2892-2923.

8 J.-F. Jhong, A. Venault, L. Liu, J. Zheng, S.-H. Chen, A. Higuchi, J. Huang, Y. Chang, *ACS Appl. Mater. Interfaces*, 2014, **6**, 9858-9870.

9 C.G. Decker, Y. Wang, S.J. Paluck, L. Shen, J.A. Loo, A.J. Levine, L.S. Miller, H.D. Maynard, *Biomaterials*, 2016, **81**, 157-168.

10 M. Kulkarni, A. O'Loughlin, R. Vazquez, K. Mashayekhi, P. Rooney, U. Greiser, E. O'Toole, T. O'Brien, M.M. Malagon, A. Pandit, *Biomaterials*, 2014, **35**, 2001-2010.

11 L.I.F. Moura, A.M.A. Dias, E. Carvalho, H.C. De Sousa, *Acta Biomaterialia*, 2013, **9**, 7093-7114.

12 S.A. Castleberry, B.D. Almquist, W. Li, T. Reis, J. Chow, S. Mayner, P.T. Hammond, *Adv Mater*, 2016, **28**, 1809-1817.

13 M. Gao, T.T. Nguyen, M.A. Suckow, W.R. Wolter, M. Gooyit, S. Mobashery, M. Chang, *Proc Natl Acad Sci USA*, 2015, **112**, 15226-15231.

14 N.J. Trengove, M.C. Stacey, S. Macauley, N. Bennett, J. Gibson, F. Burslem, G. Murphy, G. Schultz, *Wound Repair Regen*, 1999, **7**, 442-52.

15 S. Eming, H. Smola, B. Hartmann, G. Malchau, R. Wegner, T. Krieg, S. Smola-Hess, *Biomaterials*, 2008, **29**, 2932-2940.

16 B.P. Purcell, D. Lobb, M.B. Charati, S.M. Dorsey, R.J. Wade, K.N. Zellars, H. Doviak, S. Pettaway, C.B. Logdon, J.A. Shuman, P.D. Freels, J.H. Gorman, R.C. Gorman, F.G. Spinale, J.A. Burdick, *Nature Materials*, 2014, **13**, 653-661.

17 S.S. Anumolu, A.R. Menjoge, M. Deshmukh, D. Gerecke, S. Stein, J. Laskin, P.J. Sinko, *Biomaterials*, 2011, **32**, 1204-1217.

18 S. Sakai, M. Tsumura, M. Inoue, Y. Koga, K. Fukano, M. Taya, *J. Mater. Chem. B*, 2013, **1**, 5067-5075.

19 G. Tronci, H. Ajiro, S.J. Russell, D.J. Wood, M. Akashi, *Acta Biomater* 2014, **10**, 821-830.

20 Z. Fan, B. Liu, J. Wang, S. Zhang, Q. Lin, P. Gong, L. Ma, S. Yang, *Adv. Funct. Mater.*, 2014, **24**, 3933-3943.

21 M. Panca, K. Cutting, J.F. Guest, *J Wound Care*, 2013, **22**, 109-118.

22 A. Yaari, Y. Schilt, C. Tamburu, U. Raviv, O. Shoseyov, *ACS Biomater Sci Eng*, 2016, **2**, 349-360.

23 S.E.L. Bulman, P. Goswami, G. Tronci, S.J. Russell, C. Carr, *J Biomater Appl*, 2015, **29**, 1193-1200.

24 H.J. Yoo, H.D. Kim, *J Biomed Mater Res Part B: Appl Biomater*, 2008, **85B**, 326-333.

25 M.J. Waring, D. Parsons, *Biomaterials*, 2001, **22**, 903-912.

26 S.M. Bishop, M. Walker, A.A. Rogers, W.Y.J. Chen, *J Wound Care*, 2003, **12**, 125-128.

27 E.A. Rayment, T.R. Dargaville, G.K. Shooter, G.A. George, Z. Upton, *Biomaterials*, 2008, **29**, 1785-1795.

28 R.J. Wade, E.J. Bassin, C.B. Rodell, J.A. Burdick, *Nature Communications*, 2015, **6**, 6639.

29 A.H. Van Hove, M.-J.G. Beltejar, D.S.W. Benoit, *Biomaterials*, 2014, **35**, 9719-9730.

30 J.L. Leight, D.L. Alge, A.J. Maier, K.S. Anseth, *Biomaterials*, 2013, **34**, 7344-7352.

31 J.L. Leight, E.J. Tokuda, C.E. Jones, A.J. Lina, K.S. Anseth, *Proc Natl Acad Sci USA*, 2015, **112**, 5366-5371.

32 A. Francesko, D.S. Da Costa, R.L. Reis, I. Pashkuleva, T. Tzanov, *Acta Biomaterialia*, 2013, **9**, 5216-5225.

33 G. Tronci, S.J. Russell, D.J. Wood, *J Mater Chem B*, 2013, **1**, 3705-3715.

34 G. Tronci, C.A. Grant, N.H. Thomson, S.J. Russell, D.J. Wood, *J. R. Soc. Interface*, 2015, **12**, 20141079.

35 G. Tronci, A. Doyle, S.J. Russell, D.J. Wood, D.J. *Mater. Res. Soc. Symp. Proc.*, 2013, **1498**, 145-150.

36 M.T. Arafat, G. Tronci, J. Yin, D.J. Wood, S.J. Russell, *Polymer*, 2015, **77**, 102-112.

37 G. Tronci, R.S. Kanuparti, M.T. Arafat, J. Yin, D.J. Wood, S.J. Russell, *Int J Biol Macromol.*, 2015, **81**, 112-120.

38 G. Tronci, A. Doyle, S.J. Russell, D.J. Wood, *J. Mater. Chem. B*, 2013, **1**, 5478-5488.

39 A.K. Lynn, I.V. Yannas, W. Bonfield, *J Biomed Mater Res Part B: Appl Biomater,* 2004, **71B**, 343-354.

40 J. Glowacki, S. Mizuno, *Biopolymers*, 2008, **89**, 338-344.

41 G. Tronci, C.A. Grant, N.H. Thomson, S.J. Russell, D.J. Wood, *MRS Advances*, 2016, **1**, 533-538.

42 E.A. Rayment, Z. Upton, G.K. Shooter, *British Journal of Dermatology*, 2008, **158**, 951-961.

43 U. Freudenberg, A. Zieris, K. Chwalek, M.V. Tsurkan, M.F. Maitz, P. Atallah, K.R. Levental, S.A. Eming, C. Werner, *J Control Rel*, 2015, **220(A)**, 79-88.

44 J.T. Hardwicke, J. Hart, A. Bell, R. Duncan, D.W. Thomas, R. Moseley, *J Control Rel*, 2011, **152**, 411-417.

45 W.A. Bubnis, C.M. Ofner, *Anal Biochem.*, 1992, **207**, 129-133.

46 R. Jin, L.S.M. Teixeira, P.J. Dijkstra, M. Karperien, C.A. van Blitterswijk, Z.Y. Zhong, J. Feijen, *Biomaterials*, 2009, **30**, 2544-2551.

47 G. Badr, W.N. Hozzein, B.M. Badr, A.A. Ghamdi, H.M.S. Eldien, O. Garraud, *J Cell Physiol*, 2016, **231**, 2159-2171.

48 W.N. Hozzein, G. Badr, A.A.A. Ghamdi, A. Sayed, N.S. Al-Waili, O. Garraud, *Cell Physiol Biochem*, 2015, **37**, 940-954.

49 G. Badr, *Cell Physiol Biochem*, 2012, **29**, 571-582.

50 G. Badr, *Lipids in Health and Disease*, 2013, **12**, 46.

51 F. Vin, L. Teot, S. Meaume, *J Wound Care*, 2002, **11**, 335-341.

52 A. Landsman, D. Taft, K. Riemer, *Clinics in Podiatric Medicine and Surgery*, 2009, **26**, 525-533.

53 G. Tronci, A.T. Neffe, B.F. Pierce, A. Lendlein, *J Mater Chem*, 2010, **20**, 8875-8884.

54 E.R. Durham, G. Tronci, X.B. Yang, D.J. Wood, S.J. Russell, in *Biomedical Textiles for Orthopaedic and Surgical Applications*, ed. T. Blair, Elsevier; 1st Edition, 2015, 3, 45-65.

55 X. Qiao, S.J. Russell, X. Yang, G. Tronci, D.J. Wood, *J Funct Biomater*, 2015, **6**, 667-686.

56 S. Zhao, H. Wang, Y. Zhang, X. Cheng, N. Zhou, M.N. Rahaman, Z. Liu, W. Huang, C. Zhang, *Biomaterials*, 2015, **53**, 379-391.

57 Y. Loo, Y.-C. Wong, E.Z. Cai, C.-H. Ang, A. Raju, A. Lakshmanan, A.G. Koh, H.J. Zhou, T.C. Lim, S.M. Moochhala, C.A. Hauser, C.A. *Biomaterials*, 2014, **35**, 4805-4814.

58 Y. Dong, W.U. Hassan, R. Kennedy, U. Greiser, A. Pandit, Y. Garcia, W. Wang, *Acta Biomaterialia*, 2014, **10**, 2076-2085.

59 S. Majumdar, Q. Guo, M. Garza-Madrid, X. Calderon-Colon, D. Duan, P. Carbajal, O. Schein, M. Trexler, J. Elisseeff, *J Biomed Mater Res Part B: Appl Biomater*, 2016, **104B**, 300-307.

60 P.R. Stoessel, R.N. Grass, A. Sánchez-Ferrer, R. Fuhrer, T. Schweizer, R. Mezzenga, W.J. Stark, *Adv Funct Mater*, 2014, **24**, 1831-1839.

61 C. Soo, C. Rahbar, R.L. Moy, *J Dermatol Surg Oncol.*, 1993, **19**, 431-434.

62 K. Safferling, T. Sütterlin, K. Westphal, C. Ernst, K. Breuhahn, M. James, D. Jäger, N. Halama, N. Grabe, *The Journal of Cell Biology*, 2013, **203**, 691-709.